\ifdefined \singleCol
\documentclass[journal,comsoc,12pt,onecolumn,draftclsnofoot]{IEEEtran}
\else
\documentclass[journal]{IEEEtran}
\fi

\usepackage[T1]{fontenc}%
\usepackage[noadjust]{cite}

\usepackage{graphicx}
\graphicspath{{plots/pdf/},{drawings/}}
\usepackage{subfig}
\usepackage{amsmath}
\usepackage[cmintegrals]{newtxmath}
\usepackage[ruled,vlined]{algorithm2e}

\newtheorem{prop}{Proposition}

\newcommand{\myamp}{&}

\begin{document}
\title{Distributed Transmit Beamforming: Design  \\ and Demonstration from the Lab to UAVs}

\author{Samer~Hanna,~\IEEEmembership{Student~Member,~IEEE,}
        and~Danijela~Cabric,~\IEEEmembership{Fellow,~IEEE}%
\thanks{The authors are with the Electrical and Computer Engineering Department, University of California, Los Angeles, CA 90095, USA. e-mail: \mbox{samerhanna@ucla.edu}, enesk@g.ucla.edu, danijela@ee.ucla.edu.}%
\thanks{This work was supported in part by NSF  under  grant 1929874 and by the CONIX Research Center, one of six centers in JUMP, a Semiconductor Research Corporation (SRC) program sponsored by DARPA. }%
}

\maketitle

\begin{abstract}
 Cooperating radios can extend their communication range by adjusting their signals to ensure coherent combining at a destination radio. This technique is called distributed transmit beamforming. Beamforming (BF) relies on the BF radios having frequency synchronized carriers and phases adjusted for coherent combining. Both requirements are typically met by exchanging  preambles with the destination. However, since BF aims to increase the communication range, the individually transmitted preambles are typically  at low SNR and their lengths are constrained by the channel coherence time. These  noisy preambles lead to errors in frequency and phase estimation, which result in randomly changing BF gains.   To build reliable distributed BF systems, the impact of estimation errors on the BF gains need to be  considered in the design.  In this work, assuming a destination-led BF protocol and Kalman filter for frequency tracking, we optimize the number of BF radios and the preamble lengths to achieve reliable BF gain. To do that, we characterize the relations between  the BF gains distribution, the channel coherence time, and  design parameters like  the SNR, preamble lengths,   and the number of  radios.   The proposed relations are verified using simulations and via experiments using software-defined radios in a lab and  on UAVs.
\end{abstract}


\IEEEpeerreviewmaketitle

\section{Introduction}
Distributed transmit beamforming (BF) enables a group of  radios  to act as a virtual antenna array when  transmitting a common message to a destination radio. By having $N$ equal power radios beamform, the received power at the destination can increase by up to $N^2$; $N$-fold  due to transmit power increase and $N$-fold due to coherent combining~\cite{mudumbai_feasibility_2007}. The $N^2$ increase can theoretically provide  up to $N$ fold extension of communication range~\cite{mudumbai_distributed_2009}. Thus, BF  can enable long-range communications from cooperating low power devices, unable to communicate individually with a remote destination.  This can be useful for power-constrained sensor networks~\cite{jayaprakasam_distributed_2017} or UAVs  deployed in remote regions~\cite{mohanti_airbeam_2019}.

For separate radios to act as one virtual array, they need to synchronize their carrier frequencies and adjust their  phases for coherent combining at the destination. Both requirements are typically satisfied by exchanging preambles with the destination for channel phase estimation and  carrier frequency synchronization~\cite{mudumbai_distributed_2009}.  However, given that in typical BF scenarios the radios have low power and/or the destination is remote, the pre-BF SNR of individual radios is low, and there are errors in both channel estimation and destination-led frequency synchronization, which result  in phase errors in the combining signals.   These combining phase errors will lead to the BF gains  being non-deterministic and less than  $N^2$. The BF gain degradation cannot always be mitigated, especially in high mobility radios like UAV-mounted, where the channel coherence time limits the preamble lengths and makes the combining phase errors  inevitable. To build a reliable BF system despite of these  errors,  we need to specify the number of BF radios and the preamble lengths such that  a minimum desired post-BF SNR is attained with a given probability. 
Existing works have proposed many approaches  for BF  leveraging different methods for phase adjustment  and frequency synchronization~\cite{jayaprakasam_distributed_2017}. 
 Approaches for phase adjustment include explicit channel feedback from the destination~\cite{yung-szu_tu_coherent_2002},   1-bit feedback where the BF radios iteratively adjust their phase based on binary feedback from the destination~\cite{mudumbai_distributed_2010}, and roundtrip message exchange among the destination and BF radios~\cite{iii_time-slotted_2008}.  Other works proposed using the BF radios' placements to adjust the phase~\cite{hanna_feedback_2021}, however, this only works in a line-of-sight channel. For frequency synchronization, some works have relied on external frequency references like GPS~\cite{leak_distributed_2018,kramarev_event-triggered_2019}, out of band signaling~\cite{rahman_fully_2012}, and others relied on a destination preamble  along with using the extended Kalman filter (EKF) for tracking the carrier drift~\cite{quitin_demonstrating_2012}.  %
 While these works have proposed interesting approaches,  the relation between the BF gains and the pre-BF SNR, necessary for designing a reliable BF system,  was not analyzed. %

 Using the aforementioned approaches, several BF demonstrations were carried; in a controlled lab experiment, 1-bit feedback was demonstrated  using  EKF for frequency synchronization in~\cite{quitin_demonstrating_2012,quitin_scalable_2013} and out-of-band signaling in~\cite{rahman_fully_2012}. Outdoor ground based demonstrations  spanning several kilometers using explicit channel feedback were performed  in~\cite{leak_distributed_2018,kramarev_event-triggered_2019} relying on GPS for frequency synchronization. Using explicit feedback, in \cite{mohanti_airbeam_2019}, BF was demonstrated from  UAVs with the synchronization performed over wires attached to the flying UAVs.  These works have shown the potentials for BF in signal combining, yet their results  are hard to generalize to different scenarios because they are mostly empirical.

In this paper, we consider a destination-led  BF protocol using the Kalman filter (KF) for synchronization and  explicit channel feedback. For that protocol, assuming equal pre-BF SNRs,  we propose  an analytical framework relating the statistical distribution of the BF gains, with the  system  parameters including pre-BF SNR, the number of BF radios, and the duration of the exchanged preambles. Using this framework, for a given channel coherence time, we  optimize the number of radios and the length of the preambles for the BF gain to exceed a minimum SNR with a given probability, thus creating a reliable BF system. 
 To derive this framework, we derive the variance of the combining phase errors, which depends on the preamble lengths and the pre-BF SNR. Then, given the variance of the combining phase errors, we approximate the distribution of the BF gains.  The proposed framework is verified using simulations and experimentally using two BF software-defined radios (SDRs) in a lab environment. 
 To the best of our knowledge, we are the first to demonstrate fully wireless BF from flying UAVs without any wires attached.
 Our main contributions are:
 \begin{itemize}
 	\item We proposed an analytical framework describing the relations between the BF gains and the pre-BF SNR, the length of the preambles, and the number of BF radios for a destination-led BF protocol under the assumption of equal pre-BF SNRs. These relations were verified using simulations and experimentally using two BF software-defined radios.
 	\item We characterized the distribution of BF gains assuming  zero-mean normally distributed phase errors. We derived the variance of the BF gains. For large $N$, we proved the BF gain distribution approaches Gaussian and for small phase error variance  we approximated it using a Gamma distribution. 
 	\item  Using the BF framework, we proposed an approach to determine the minimum number of BF radios and the shortest BF preambles to meet a minimum post-BF SNR with a given probability and verified that it meets the requirements using simulations.
 \end{itemize}

\renewcommand{\b}[1]{\boldsymbol{\mathrm{#1}}}
\providecommand{\h}[1]{\ensuremath{\b{h}_{#1}}}

\newcommand{\C}[1]{\mathbb{C}^{#1}}
\newcommand{\R}[1]{\mathbb{R}^{#1}}
\newcommand{\Z}{\mathbb{Z}}
\newcommand{\U}{\mathcal{U}}
\newcommand{\mI}{\b{I}}
\newcommand{\E}[1]{\mathbb{E}\{#1\}}

\newcommand{\mfh}[1]{ f_{#1}}
\newcommand{\mph}[1]{\phi_{#1}}
\newcommand{\mah}[1]{a_{#1}}

\newcommand{\mpwd}[1]{\hat{\phi}'_{#1}}
\newcommand{\mpw}[1]{\hat{\phi}_{#1}}
\newcommand{\mfw}[1]{\hat{f}_{#1}}

\newcommand{\mSNRpre}{\gamma_{\text{preBF}}}
\newcommand{\mSNRpost}{\gamma_{\text{postBF}}}
\newcommand{\mSNRms}{\gamma_{\text{DR}}}
\newcommand{\mSNRmin}{\gamma_{\text{min}}}

\newcommand{\mtsyn}{t_{\text{syn}}}
\newcommand{\mtph}{t_{\text{ph}}}
\newcommand{\mtfb}{t_{\text{fb}}}
\newcommand{\mtgI}{t_{\text{g1}}}
\newcommand{\mtgII}{t_{\text{g2}}}
\newcommand{\mtgIII}{t_{\text{g3}}}
\newcommand{\mtov}{t_{\text{ov}}}
\newcommand{\mtp}{t_{\text{p}}}

\newcommand{\mnsyn}{N_{\text{syn}}}
\newcommand{\mnph}{N_{\text{ph}}}
\newcommand{\mnfb}{N_{\text{fb}}}
\newcommand{\mngI}{N_{\text{g1}}}
\newcommand{\mngII}{N_{\text{g2}}}
\newcommand{\mngIII}{N_{\text{g3}}}
\newcommand{\mlzc}{M}
\newcommand{\mnzc}{N_{\text{ZC}}}
\newcommand{\mnov}{N_{\text{ov}}}

\newcommand{\mte}{t_e}
\newcommand{\mpe}[1]{\phi^e_{#1}}

\newcommand{\mTs}{T_s}

\newcommand{\mvar}[1]{\text{var}\{#1\}}
\newcommand{\mcov}[1]{\text{cov}\{#1\}}
\newcommand{\mvarN}{\sigma^2}
\newcommand{\mvarPe}{\sigma^2_{e}}
\newcommand{\mstdPe}{\sigma_{e}}
\newcommand{\mvarF}{\sigma^2_{f}}
\newcommand{\mvarPh}{\sigma^2_{ph}}
\newcommand{\mvarPhe}{\sigma^2_{phe}}
\newcommand{\mvarFb}{\sigma^2_{fb}}
\newcommand{\mvarFbe}{\sigma^2_{fbe}}
\newcommand{\mPdt}{P^{D}_{T}}

\newcommand{\mpout}{p_{\text{out}}}
\section{System Model and Distributed BF Protocol}
\subsection{System Model}
Consider $N$ identical radios collaborating to beamform a  message  to  a destination radio $D$ in a narrowband flat-fading channel. The BF radios can be remotely deployed Internet-of-Things devices communicating with a gateway or UAVs  communicating with a ground station.  The message is encoded in the complex baseband signal $m(t)$, which is assumed to have unity power. The   $n$-th radio transmits a signal  $z_n(t)$ and the combined baseband signal at the destination is given by 
\begin{equation}
	\label{eq:rec_sig}
	y(t)=  \sum_{n=1}^{N}  \mah{n} z_n(t)  \exp \{ j(2\pi \mfh{n} t + \mph{n} )\} + w(t)
\end{equation}
where between the destination and the $n$-th radio, $\mah{n}$ is the channel amplitude, $\mfh{n}$ is the  carrier frequency offset, and $\mph{n}$ is the phase offset. The white Gaussian noise process is given by $w(t)$ and has power spectral density $N_0/2$. The phase and frequency offsets result from the lack of synchronization between the local oscillators of the radios, the wireless propagation environment, and the Doppler frequency offsets resulting from the relative motion of radios. While these phenomena make the phase and frequency offsets time varying,  we assume that the message is shorter than the resulting channel coherence time and we approximate them as constant for one message. 

For the signal $m(t)$ to combine coherently at the destination, the BF radios  need to compensate for the phase  and frequency offsets before transmission. The compensated signal transmitted by radio $n$, thus, is given by 
\begin{equation}
	z_n(t)= m(t)  \exp \{-j(2\pi \mfw{n} t + \mpw{n} )\}
\end{equation}
where $\mfw{n}$ and $\mpw{n}$ are the $n$-th radio estimates of the frequency and phase offsets obtained through the BF protocol, which is described later.    The received signal  can be rewritten as
\begin{equation}
	y(t)=  m(t) \sum_{n=1}^{N}  \mah{n}   \exp \{ j \mpe{n}(t) \} + w(t)
\end{equation}
where the combining phase error from radio $n$ at instant $t$ is given by 
\begin{equation}
    \label{eq:phase_error}
	\mpe{n}(t) = (2\pi (\mfh{n}-\mfw{n}) t + (\mph{n}-\mpw{n} ) )
\end{equation}
Due to residual frequency errors, the combining phase error increases with time. However, we are only interested in evaluating the BF gain during the payload. Considering the evaluation instance to be  $\mte$ seconds after the phase estimation, we get $\mpe{n}=\mpe{n}(\mte)$.  The beamforming gain at instant $\mte$ can be defined as the ratio between the energy of the combined signals to that of the  individual transmissions
\begin{equation}
	\label{eq:g_channel}
	G=\frac{ \lVert \sum_{n=1}^{N} \mah{n} \exp \{ j \mpe{n} \}\rVert^2} { \sum_{n=1}^{N} \mah{n}^2 }
\end{equation}
Each BF radio is assumed to transmit at its maximum power level $P_T$, which is common to all radios.  We also assume that the BF radios are deployed in proximity from each other far from the destination, and hence they experience similar signal attenuation.  Given these assumptions, we get $\mah{n}=a$ for all $n$, where $a$ is the path loss.  In that case,  $G$ simplifies to~\cite{richard_brown_receiver-coordinated_2012}
\begin{equation}
	G=\frac{1} { N }  \left| \sum_{n=1}^{N}  \exp \{ j \mpe{n} \}\right|^2
\end{equation}
The pre-BF SNR at the destination from one radio  is given by 
\begin{equation}
	\mSNRpre=\frac{a^2 P_T}{N_0} 
\end{equation}
and the post-BF SNR of the combined signal from all $N$ BF radios is equal to 
\begin{equation}
	\label{eq:postBF_SNR}
	\mSNRpost = N\mSNRpre  G 
\end{equation}
The signals transmitted by the destination to the BF radios experience an SNR  given by
\begin{equation}
	\mSNRms=\frac{a^2 \mPdt}{N_0}
\end{equation}
where the destination has a  transmit power $\mPdt$. The destination   transmit power is assumed to be equal to or larger than that of the BF radios, that is $\mPdt\geq P_T$. Note that the post-BF SNR follows the same distribution of $G$, which we need to know to realize a minimum post-BF SNR with a given probability. As for $G$, it depends on $\mpe{n}$, which results from the estimation errors during the BF protocol.

\subsection{Beamforming Protocol}
\begin{figure}[t!]
	\centering
	\includegraphics{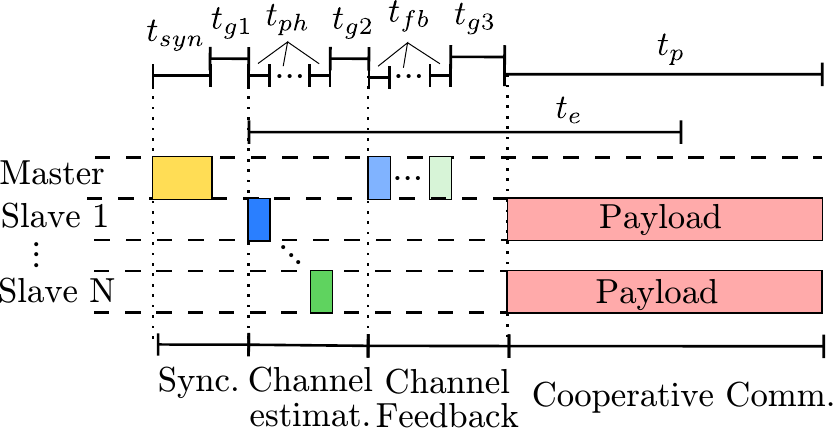}
	\caption{Timing diagram of BF protocol. The destination is the master and BF radios are the slaves. }	
	\label{fig:timing_diagram}
\end{figure}
We start by describing the BF protocol, which aims to provide each BF radio with  estimates of its phase and frequency offsets $\mpw{n}$ and $\mfw{n}$.  We consider a master-slave beamforming protocol ; the destination radio is used as a master since it has a larger transmit power  and the slaves are the beamforming radios.  The master initiates the beamforming procedure and sends a preamble for frequency synchronization.  After correcting their frequencies, the slaves send a channel estimation preamble to the master. The master  calculates a phase estimate $\mpwd{n}$  and transmits it back to the slaves that receive a slightly different value $\mpw{n}$ due to feedback errors. Once each slave knows $\mpw{n}$ and $\mfw{n}$, they can start transmitting their payload.

 In Fig.~\ref{fig:timing_diagram}, we illustrate the transmitted signals.  All the signaling is performed on the same frequency band, hence, all transmissions are received by all radios.  The different beamforming stages can be described as follows
\begin{enumerate}
	\item Synchronization: The master sends a synchronization preamble of duration $\mtsyn$. Using this signal each slave estimates  its frequency offset $\mfw{n}$. This preamble is also used as a time reference.  A guard time  of duration $\mtgI$ is provided for the slaves to process the signals.
	\item Channel Estimation: Each slave during an allocated time slot sends a channel estimation preamble of duration $\mtph$. The master estimates $\mpwd{n}$ from each slave. A guard time  $\mtgII$ is used.
	\item Channel Feedback: The master sends  $\mpwd{n}$ back to the slaves and due to feedback error each slave receives a slightly different phase estimate $\mpw{n}$. A guard time  $\mtgIII$ is used.
	\item Cooperative Communication: After estimating  $\mfw{n}$ and receiving  $\mpw{n}$, all slaves adjust their signals accordingly and transmit their payload of duration $\mtp$. 
\end{enumerate}
The duration of the BF   overheads incurred by the protocol is given by
\begin{equation}
	\mtov = \mtsyn + N (\mtph + \mtfb) + \mtgI +\mtgII + \mtgIII 
\end{equation}
 All the signal processing is assumed to be done in discrete time domain, hence all the time durations are assumed to be an integer multiple of the sampling time $\mTs$. The time overhead can be written in terms of samples as 
\begin{equation}
	\label{eq:overhead_n}
	\mnov = \mnsyn + N (\mnph + \mnfb) + \mngI +\mngII + \mngIII 
\end{equation}
where  $\mnov$ is defined as $\mnov=\mtov/\mTs$ and the remaining number of samples are defined similarly.  As we can see from~(\ref{eq:overhead_n}), the beamforming overheads scale with the number of BF radios $N$. For short coherence time channels, the overheads $\mnov$ are typically constrained, and to increase $N$ while keeping $\mnov$  constant, the duration of the preambles needs to be reduced. 
 Note that we  assume that the payload is already shared among all the slaves. This can be achieved using a network broadcasting protocol~\cite{williams_comparison_2002}, which we do not discuss in this work. As for the guard time, it is dependent on the implementation of the BF protocol. A more optimized implementation using an FPGA for instance would require shorter guard times than an implementation using a general purpose processor. Also, note that  cooperative communication only  requires the BF radios to be synchronized with each other and not necessarily with the destination. However, in order to use channel estimates from the  destination, they need to be synchronized with the destination. 

  Since BF is used to improve the SNR where the individual pre-BF SNR is low,  the estimation errors  within the protocol can not be neglected and will lead to a combining phase error $\mpe{n}$ as given by (\ref{eq:phase_error}). At the evaluation time $\mte$, the variance of the combining phase error $\mvarPe$ defined as $\mvar{\mpe{n}}$ is given by
\begin{equation}
	\label{eq:var_pe}
	\mvarPe=  (2\pi \mte)^2 \mvarF +  \mvarPh + \mvarFb 
\end{equation}
where the  frequency estimation  variance is given by $\mvarF=\mvar{\mfh{n}-\mfw{n}}$, the phase estimation and feedback variances are given by $\mvarPh=\mvar{\mph{n}-\mpwd{n}}$ and $\mvarFb=\mvar{\mpwd{n}-\mpw{n}}$ respectively.

In the following Sections,~\ref{sec:freq} and~\ref{sec:phase},  we discuss the waveforms and estimators used for frequency estimation and phase estimation \& feedback respectively. We provide expressions for their error variances in terms of the pre-BF SNR and the preamble lengths. We argue that the resulting phase errors follow a zero-mean Gaussian distribution.  For zero-mean Gaussian distributed phase errors with variance $\mvarPe$, we approximate the distribution of the BF gain in Section~\ref{sec:bf_gain_analysis} to complete the BF framework. This framework is numerically and experimentally verified in Section~\ref{sec:validation}. After verifying the framework, we show how  it can be used for designing BF systems in Section~\ref{sec:bf_design}. The BF design procedures are illustrated using example scenarios in Section~\ref{sec:scenarios}.

\newcommand{\mfEst}{\eta_f}

\newcommand{\myf}{y_f}
\newcommand{\mxf}{x_f}
\newcommand{\mvarFe}{\sigma^2_{fe}}
\newcommand{\mvarFk}{\sigma^2_{fk}}
\newcommand{\mtcycle}{t_{\text{cyc}}}
\section{Frequency Synchronization}
\label{sec:freq}
The objective of frequency synchronization is to eliminate the frequency offset between the destination and the BF radios. We start by discussing the signals used for synchronization and the proposed oneshot estimator and its variance. Then we discuss frequency tracking using Kalman filter assuming multiple successive BF cycles.%

\subsection{Frequency Offset Estimation}
\label{subsec:freq_est}
For frequency synchronization, we use a preamble consisting of $\mnzc$ repetition of a Zadoff-Chu (ZC) sequence of length $\mlzc$ similar to~\cite{yan_aeroconf_2019}, satisfying $\mnsyn=\mnzc \mlzc$. The frequency estimator   calculates the auto-correlation statistic
\begin{equation}
	\label{eq:freq_est_statistic}
	\mfEst = \sum_{k=0}^{(\mnzc-)M-1}  \myf^*[k] \myf[k+M]
\end{equation}
where $\myf[k]$ is the noisy received preamble with the frequency offset, and  $()^*$ denotes the conjugate operator.  The frequency offset estimate at slave $n$ is thus given by 
$
	\mfw{n}=\frac{1}{2\pi \mTs \mlzc}\angle \mfEst
$
where $\angle(\cdot)$ denotes the phase of a complex number calculated using arctan. The term $\angle \mfEst$ calculates the phase difference between two successive sequence repetitions, under the assumption that $\mlzc$  is small such that no phase wrapping occurs.  The error variance for this  estimator is given by~\cite[eq.70]{lank_semicoherent_1973} 
\begin{equation}
	\label{eq:freq_est_var}
	\mvarFe= \left(\frac{1}{\mlzc(\mnzc-1)^{2} \mSNRms}+\frac{1}{2 \mlzc(\mnzc-1)\mSNRms^{2}}\right)\frac{1}{(2\pi \mlzc \mTs)^{2}}
\end{equation}
This estimator is unbiased thus  $\E{\mfw{n} -\mfh{n} }=0$  and was derived using a linear approximation of the arctan assuming $\mfEst$ has a high SNR. By choosing $\mlzc$ to be large, using the central limit theorem, the  distribution of $\mfEst$ can be approximated by Gaussian, thus making $\mfw{n} -\mfh{n}$, which is approximated as linear in  $\mfEst$,  a zero mean Gaussian RV. However,  at low SNR of $\mfEst$,  $\angle \mfEst$ becomes uniform and the expression of $\mvarFe$  no longer applies. This regime can be avoided by increasing $\mnzc$, otherwise, the BF gains will be too low to be of practical importance. Note that in practice the frequency offset is correlated among  successive packets with short separation. This estimator, referred to as a oneshot frequency estimator, does not benefit from this correlation.

\begin{table}[t]
	\renewcommand{\arraystretch}{1.5}
	\caption{ Kalman Filter relations \label{tbl:KF}}
	\centering
	\ifdefined \singleCol 
		\begin{tabular}{|c|c|c|}
			\hline
			Model & Update & Predict \\	\hline
			\parbox{2.0 in}{
			\begin{equation}
				\label{eq:freq_wiener_disc}
				x_{k}=x_{k-1}+w_{k-1}
			\end{equation}
			\begin{equation}
				\label{eq:freq_measur}
				z_{k}=x_{k}+v_{k}
		\end{equation} }
		& 
		\parbox{2.0 in}{
			\vspace{0.1cm}
		\begin{equation}
			\label{eq:kf_gain}
			K_k=\frac{p_{k|k-1}}{p_{k|k-1}+r}
		\end{equation}
		\begin{equation}
			\label{eq:kf_current}
			x_{k|k}=x_{k|k-1}+K_{k} (z_{k}-x_{k|k-1})
		\end{equation}
		\begin{equation}
			p_{k|k}=(1-K_{k})p_{k|k-1}
		\end{equation}
		}
	 & 
	 \parbox{2.0 in}{
	\begin{equation}
	 \label{eq:kf_udpate}
	 x_{k+1|k}=x_{k|k}
	\end{equation}
	\begin{equation}
	\label{eq:kf_var_udpate}
	p_{k+1|k}=p_{k|k}+q
	\end{equation}
	}
	 \\	\hline
	\end{tabular} 
	\else
	
		\begin{tabular}{|c|}
			\hline
			Model \\	\hline
			\parbox{2.0 in}{
				\begin{equation}
					\label{eq:freq_wiener_disc}
					x_{k}=x_{k-1}+w_{k-1}
				\end{equation}
				\begin{equation}
					\label{eq:freq_measur}
					z_{k}=x_{k}+v_{k}
			\end{equation} }
			  \\ \hline  Update \\	\hline 
			\parbox{2.0 in}{
				\vspace{0.1cm}
				\begin{equation}
					\label{eq:kf_gain}
					K_k=\frac{p_{k|k-1}}{p_{k|k-1}+r}
				\end{equation}
				\begin{equation}
					\label{eq:kf_current}
					x_{k|k}=x_{k|k-1}+K_{k} (z_{k}-x_{k|k-1})
				\end{equation}
				\begin{equation}
					p_{k|k}=(1-K_{k})p_{k|k-1}
				\end{equation}
			}
			\\ \hline Predict \\	\hline 
			\parbox{2.0 in}{
				\begin{equation}
					\label{eq:kf_udpate}
					x_{k+1|k}=x_{k|k}
				\end{equation}
				\begin{equation}
					\label{eq:kf_var_udpate}
					p_{k+1|k}=p_{k|k}+q
				\end{equation}
			}
			\\	\hline
		\end{tabular} 
	
	\fi
\end{table}
\newcommand{\mxcf}{X_f}
\newcommand{\mxcp}{X_p}
\newcommand{\mstdcf}{\sigma_2}
\newcommand{\mstdcp}{\sigma_1}
\newcommand{\mwinf}{W_2}
\newcommand{\mwinp}{W_1}
\newcommand{\mcoht}{\tau_c}
\newcommand{\mcohl}{R_L}

\subsection{Interpacket Frequency Tracking using Kalman Filter}
If beamforming is performed periodically at a fixed cycle duration $\mtcycle$ shorter than the channel coherence time, the frequency estimates between packets  at each slave are correlated.  Kalman filter (KF), thus, can be used to track the frequency to reduce the estimation variance. 
The drift system model and the KF equations are given in Table~\ref{tbl:KF} for one BF radio following the conventional KF notation~\cite{thacker_tutorial_1998}.   The frequency process drift  and measurement models are given by (\ref{eq:freq_wiener_disc}) and (\ref{eq:freq_measur}), respectively, where $x_{k}$ is the true frequency value in Hz (previously denoted by $\mfh{n}$) and $z_{k}$ is the measured frequency at time $k \mtcycle$. The noise terms for the process  $w_{k}$ and the measurement $v_{k}$ are assumed to be zero mean Gaussian RV and their  variances are $q$ and $r$ respectively.
 For the KF update equations, at step $k$, $K_k$ is the Kalman gain, $x_{k|k-1}$ is the prediction of $x$  and $p_{k|k-1}$ is the error variance given $z_{k-1}$. The value of $x_{k|k}$ is the predicted frequency offset    and $p_{k|k}$ is its error variance given  $z_{k}$.

By  substituting (\ref{eq:kf_gain}) in (\ref{eq:kf_current}) and using (\ref{eq:kf_udpate})  we get 
\begin{equation}
	\label{eq:kf_simple}
	x_{k|k}=\frac{r}{p_{k|k-1}+r}x_{k-1|k-1}+\frac{p_{k|k-1}}{p_{k|k-1}+r}z_{k}
\end{equation}
from which we can see that the KF creates a weighted average between the previous prediction and the current measurement. The weights of this average are based on the predicted process variance $p_{k|k-1}$ and the measurement variance $r$. The larger the process variance relative to the measurement variance,  the more  weight is given to the measured value and vice versa. Since~(\ref{eq:kf_simple}) is a linear equation, if $z_{k}$ is a zero mean Gaussian RV, the output of KF will also be zero-mean and Gaussian.
 For  BF, we are interested in calculating the KF error variance.
\begin{prop}
	\label{prop:kf_var}
	The steady state  frequency estimation error variance of  KF from Table~\ref{tbl:KF} is
	\begin{equation}
		\label{eq:kf_op_var}
		\mvarFk=\frac{-q+q\sqrt{1+4\frac{r}{q}}}{2}
	\end{equation}
\end{prop}
The proof is in  Appendix~\ref{ap:kf_var}. Using (\ref{eq:kf_op_var}) and assuming the system variances are accurately known, we argue that KF never increases the error variance.
 By rewriting (\ref{eq:kf_op_var}), as $\mvarFk=\frac{-q+\sqrt{q^{2}+4qr}}{2}$, we can see that $\mvarFk$ is non-decreasing in $q$ and if $q=0$, at convergence the error variance $\mvarFk=0$ for any $r$. For  $q>>r$, $r/q$ is small and  using the approximation $\sqrt{1+4\frac{r}{q}}\approx 1 + 2\frac{r}{q}$, we get $\mvarFk=r$. Thus if $q$ and $r$ are perfectly known, the error variance reduction due to KF is higher for large $r/q$ and, in the worst case scenario for small $r/q$, KF will give the measurement variance   $\mvarFk=r$,  as if we did not use KF. However, if the values of $q$ and $r$ used in KF do not match the system, this result does not hold and KF might deteriorate the frequency estimation.
 Note that  the extended KF (EKF) can  track both phase and frequency and might yield a smaller variance than  KF which only tracks the frequency. However, EKF can diverge due to phase wrapping~\cite{quitin_scalable_2013}, which is not desirable in a reliable BF system, and thus was not considered in this work. %

\newcommand{\myph}{y_\text{ph}}
\newcommand{\mxph}{x_\text{ph}}
\newcommand{\myfb}{y_\text{fb}}
\newcommand{\mxfb}{x_\text{fb}}
\newcommand{\mxfbpvec}{\b{x}_\text{fbp}}
\newcommand{\mxfbvec}{\b{x}_\text{fb}}

\newcommand{\mphEst}{\eta_{\text{ph}}}
\newcommand{\mfbEst}{\eta_{\text{fb}}}
\section{Phase Estimation and Feedback}
\label{sec:phase}
The objective of the phase estimation and feedback is to have the slaves modify their signals to ensure coherent combining at the destination. In the phase estimation stage, each slave transmits a known signal $\mxph[n]$ consisting of $\mnph$ samples. The master receives the noisy signal $\myph[k]$. The proposed estimator calculates the correlation
$
\mphEst=\sum_{k=0}^{\mnph-1} \mxph[k] \myph[k]
$,
from which the phase estimate is calculated using
$
	\mpwd{n}=\angle \mphEst
$.
The variance of this estimator is given by~\cite{tretter_estimating_1985}
\begin{equation}
	\label{eq:ph_est_var}
	\mvarPhe=\frac{1}{2 \mnph \mSNRpre }
\end{equation}
where $\mnph \mSNRpre $ is the  SNR of $\mphEst$. The phase error $\mpwd{n}$ follows a zero mean Gaussian distribution as long as the  SNR of $\mphEst>>1$ ~\cite{tretter_estimating_1985}, which is the regime of interest.

As for the phase feedback, we use in-band feedback where the value of $\mpwd{n}$ is encoded in the phase difference between two identical preambles to counter hardware phase ambiguity. Let the phase feedback preamble be given as a vector $\mxfbpvec \in \C{\mnfb}$. The master transmits the sequence
\begin{equation}
	\mxfbvec = [\mxfbpvec^T \ \ \mxfbpvec^T e^{j\mpwd{1}}  \cdots \mxfbpvec^T e^{j\mpwd{n}}  \cdots  \mxfbpvec^T e^{j\mpwd{N}}]^T
\end{equation}
Once received as $\myfb[k]$ with added noise, slave $n$ estimates the phase difference between the first preamble and the $n$-th preamble using the statistic $\mfbEst=\sum_{k=0}^{\mnfb-1} \myfb[k] \myfb[k+n \mnfb]$ and calculates the angle $\mpw{n}=\angle \mfbEst$. The variance of the feedback is similar to that used for frequency estimation in (\ref{eq:freq_est_var}) (with $\mnzc=2$, $\mlzc=\mnfb$) and is given by
\begin{equation}
	\label{eq:fb_est_var}
	\mvarFbe= \left(\frac{1}{\mnfb \mSNRms}+\frac{1}{2 \mnfb \mSNRms^{2}}\right)
\end{equation}
Note that there are other ways to feedback the phase estimates, however, this approach is simple to implement. Another alternative was to encode the values of $\mpwd{n}$ as floating-point numbers and transmit them using digital modulation. However, since we are considering a low SNR  and a mistake in one of the most significant bits can be detrimental, we would need to implement channel coding. This would add unnecessary complexity to our protocol.%

\section{Beamforming Gain Analysis}
\label{sec:bf_gain_analysis}
\begin{figure}[t!]
	\centering
	\subfloat[The mean BF Gain and its  standard deviation as error bars.]{\includegraphics{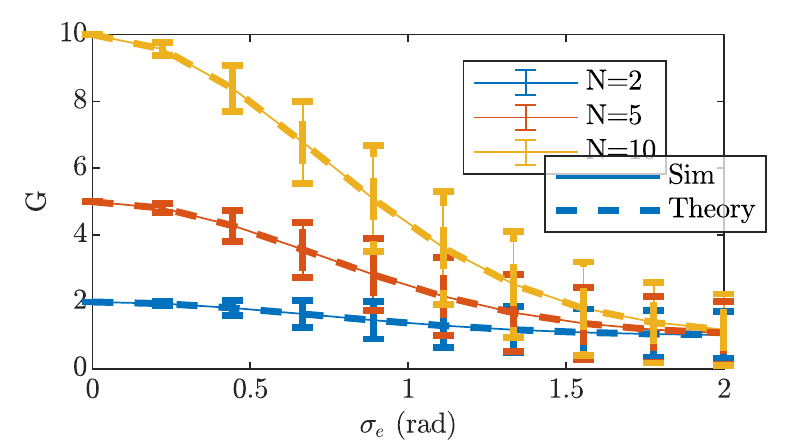}\label{fig:bf_gain}}
	\ifdefined \singleCol\else	\\	\fi
	\subfloat[The relation between the BF variance and $N$ for fixed $\mvarPe$.]{\includegraphics{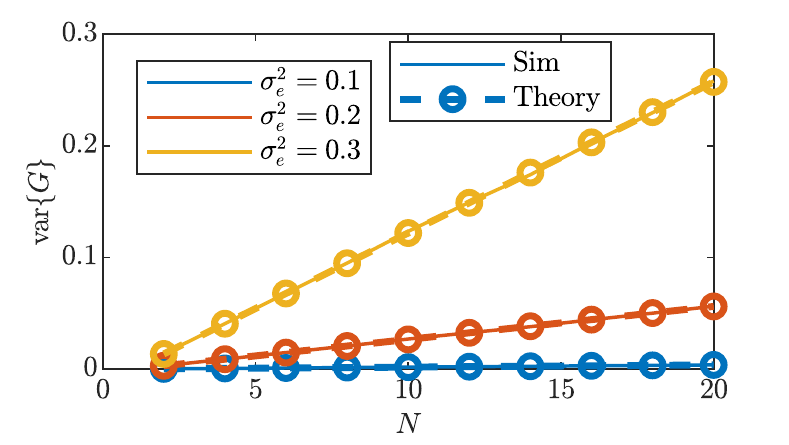}\label{fig:bf_gain_var}}
	\caption{The relation between BF gain, $N$, and $\mvarPe$.}
	\label{fig:bf_gv}
\end{figure}

\begin{figure}[t!]
	\centering
	\subfloat[The distribution of $G$ for small $N$ and small $\mvarPe=0.1$.]{\includegraphics{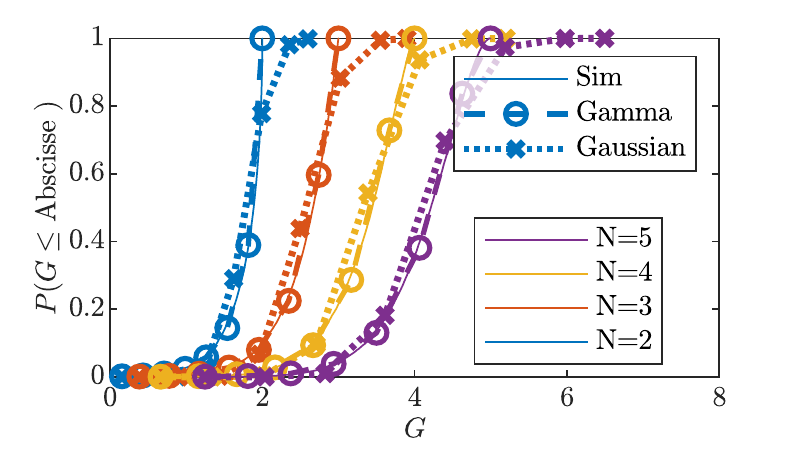}\label{fig:bf_dist_n_2}}
	\ifdefined \singleCol\else	\\	\fi
	\subfloat[The distribution of $G$ for large $N$ and large $\mvarPe=1$.]{\includegraphics{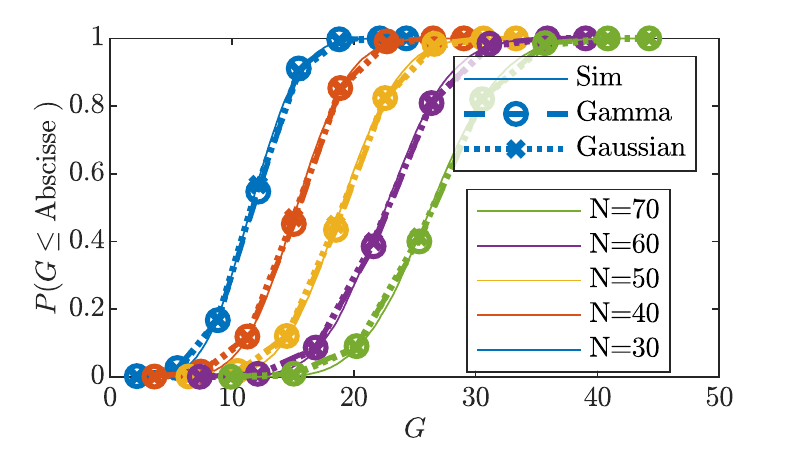}\label{fig:bf_dist_n_30}}
	\caption{The distribution of $G$ for different $N$ and $\mvarPe$}
	\label{fig:bf_dist}
\end{figure}

In this section, our objective is to approximate the distribution of $G$,  assuming that the $\mpe{n}$ are independent Gaussian random variables (RVs) with zero mean and variance $\mvarPe$.   The Gaussian assumption applies to our protocol because the errors of the proposed estimators are independent and  can be approximated by a zero-mean Guassian RVs. Hence, their sum according to (\ref{eq:phase_error}) is also zero-mean Gaussian.  We start by calculating the mean and variance of the distribution.
\begin{prop}
	\label{prop:G_mean_var}
	For signals combining from $N$ radios having independent zero  mean Gaussian phase with variance $\mvarPe$, the mean and the variance of the BF gains $G$ are given by 
	\begin{equation}
	\label{eq:bf_gain_mn}
	\E{G}=1+(N-1)e^{-\mvarPe}
	\end{equation}
	\begin{equation}
	\label{eq:bf_gain_var}
	\mvar{G}= \frac{(N-1)}{N}(1-e^{-\mvarPe})^{2}\left((1-e^{-\mvarPe})^{2}+2Ne^{-\mvarPe})\right)
	\end{equation}
\end{prop}
The proof is in Appendix~\ref{ap:G_mean_var}.  Note that the mean was previously derived in~\cite{richard_brown_receiver-coordinated_2012}.
In Fig.~\ref{fig:bf_gain}, we plot the average BF gain using (\ref{eq:bf_gain_mn}) as a function of $\mstdPe$ with the error bars representing the standard deviation ($\sqrt{\mvar{G}}$). For $\mstdPe=0$, we get a BF gain of $N$ as we ideally expect. As  $\mstdPe$ increases, the mean BF gains  decrease and their variances increase  and this happens faster for larger $N$. Thus when designing a BF system unless $N$ and  $\mstdPe$ are small, we can not assume a perfect $N$ fold power increase due to BF.  To verify the derived mean and variance, for each value of $N$ and $\mstdPe$, we sampled  100,000 zero mean Gaussian RVs of variance $\mvarPe$ for each radio and added them to calculate $G$ numerically. The simulations shown in~Fig.\ref{fig:bf_gain} as thick dashed lines with dashed error bars overlap the derived expressions verifying Proposition~\ref{prop:G_mean_var}. 

To better understand the variance behavior with $N$, for small $\mvarPe$,  we simplify (\ref{eq:bf_gain_var}) to get  $\mvar{G}\approx 2Ne^{-\mvarPe }(1-e^{-\mvarPe})^{2}$. Thus the variance increases linearly with the number of slaves for small $\mvarPe$. The linear relation between   $\mvar{G}$ and $N$ is illustrated in~Fig.\ref{fig:bf_gain_var}. The higher the value of $\mvarPe$, the larger the slope. The large discrepancy in the values of the variance with $N$ shows the importance of considering the distribution of $G$ and not just its mean  in the design of reliable BF systems. Next, we approximate the distribution of $G$. First, we consider the case of large $N$ using the central limit theorem. Then,  we consider the case for a small $N$ and small $\mvarPe$ and use the Taylor series to derive the approximation. 
\begin{prop}
	\label{prop:g_gaussian}
	For large $N$, the distribution of G tends to a Gaussian distribution with mean and variance given by Proposition~\ref{prop:G_mean_var}.
\end{prop}
\newcommand{\mXX}{\b{X}}
\newcommand{\mX}[1]{\mXX_{#1}}

\newcommand{\mgammarv}{X_{\gamma}}
\begin{prop}
	\label{prop:g_gamma}
	For small combined phase error variance $\mvarPe$ or for large $N$, the distribution of $G$ can be approximated by $N-\mgammarv$ where $\mgammarv$ is a random variable following the Gamma distribution $\mgammarv \sim \Gamma(K,\theta)$ with
	\begin{equation}
	\label{eq:g_gamma_k}
	K=\frac{N(N-1)}{(1-e^{-\mvarPe})^{2}+2Ne^{-\mvarPe}}
	\end{equation}
	\begin{equation}
	\label{eq:g_gamma_theta}
	\theta=\frac{1}{N} (1-e^{-\mvarPe})\left((1-e^{-\mvarPe})^{2}+2Ne^{-\mvarPe})\right)
	\end{equation}
\end{prop}
The proofs are in  Appendices~\ref{ap:prop_g_normal} and~\ref{ap:g_gamma} respectively. We start by plotting the empirical cumulative distribution function (CDF) of $G$ for  small $N$ and a small $\mstdPe=0.1$  in Fig.~\ref{fig:bf_dist_n_2}. We can see that the distribution is not Gaussian and is accurately approximated by the Gamma distribution. Then, we consider a large $N\geq 30$ and relatively large value of $\mstdPe=1$ in Fig.~\ref{fig:bf_dist_n_30}. From that Figure, we can see that all three CDFs overlap for large $N$ and large $\mstdPe$ verifying Prop.~\ref{prop:g_gaussian} and~\ref{prop:g_gamma}. Based on these results, since the Gamma distribution applies to a wider range of $N$ and $\mvarPe$, we use it later to approximate the BF gain distribution.  Note that neither approximation is accurate for small values of $N$ and a large value of $\mvarPe$, however, in this regime the BF gains are small with a large variance, which  is not of practical importance.   It is  important to note that the derived variance and distribution approximation in this section apply to any BF protocol where the phase error $\mpe{n}$ is independent for all $n$ and can be approximated by zero-mean Gaussian RVs.  For our protocol, the value of $\mvarPe$ can depend on $N$ for scenarios where the BF overhead  $\mnov$ is  constrained by the channel coherence time. In such scenarios, the duration of each preamble decreases as $N$ increases to satisfy the fixed $\mnov$.  Thus the estimators error variances and consequently  $\mvarPe$ increase with $N$. The dependence between $N$ and $\mvarPe$ is considered when designing the BF preambles in short coherence channels later in Section~\ref{sec:scenarios_small_uav}.

\section{Numerical and Experimental Validation}
\label{sec:validation}
In this Section, after deriving the BF framework, we verify it numerically and experimentally and we show that it can be used to  predict the BF gains at different SNRs. Using UAV experiments and emulation over a UAV channel trace, we evaluate the impact of the channel coherence time on the BF gains.
\begin{table}[t]
	\renewcommand{\arraystretch}{1.5}
	\caption{Beamforming Waveform Specifications~\label{tbl:timing}}
	\centering
	\begin{tabular}{|c|p{
				\ifdefined \singleCol 3.5in \else 2.4 in	\fi	
			}|}
		\hline
		Scenario & Parameters \\	\hline
		Simulation &  $\mnzc=10$, $\mlzc=63$, $\mtsyn=0.63 ms$, $\mtph=0.1 ms$,
		$\mtfb=0.1 ms$,  $\mtgI=\mtgII=\mtgIII=1 ms$, $\mtp=12 ms$, $\mte=9 ms$, $\mtcycle=50ms$ \\ \hline
		Lab &  $\mnzc=10$, $\mlzc=63$, $\mtsyn=0.63 ms$, $\mtph=0.1 ms$,
		$\mtfb=0.1 ms$,  $\mtgI=6ms$ $\mtgII=4ms$ $\mtgIII=16 ms$, $\mtp=10 ms$, $\mtcycle=180ms$ \\ \hline
		UAV &  $\mnzc=10$, $\mlzc=63$, $\mtsyn=0.63 ms$, $\mtph=0.1 ms$,
		$\mtfb=0.1 ms$,  $\mtgI=6ms$ $\mtgII=4ms$ $\mtgIII=11 ms$, $\mtp=1 ms$, $\mtcycle=75ms$ \\ \hline
	\end{tabular} 
\end{table}

\subsection{Numerical Validation}

We simulated the BF protocol between a destination radio  and $N$ BF radios. During a BF cycle,   signals transmitted from BF radio $n$ to the destination is multiplied by $e^{ j(2\pi \mfh{n} t + \mph{n})}$ with noise added to realize the SNR $\mSNRpre$. Any signal transmitted the other way  uses the negative value of $\mfh{n}$ with noise added to realize the SNR $\mSNRms$ . At the start of each BF cycle, for BF slave $n$, we sample uniform  random phase $\mph{n}$ and $\mfh{n}$ is generated using a discrete Wiener process as described in (\ref{eq:freq_wiener_disc}) having variance $q$. Since we are assuming that the signal is transmitted within the channel coherence time, both frequency and phase are assumed to be constant during the same BF cycle.

The signals transmitted follow the BF protocol. For phase estimation and feedback, we used the estimators discussed in Section~\ref{sec:phase} and for frequency offset we either used the oneshot estimator from Section~\ref{subsec:freq_est} alone or combined with  KF. To avoid errors in  measuring the BF gain, the combined signal magnitude was evaluated  at  time  $\mte$  before adding the noise.

In our simulations, we considered $N=5$ BF radios using a sampling rate of 1MHz ($\mTs=1 \mu s$). The exact duration of each preamble is given in the first row of Table~\ref{tbl:timing} and we used $q=0.18$.%
The evaluation time $\mte=9 ms$  is in the middle of the payload. One million BF cycles were simulated.

\begin{figure}[t!]
	\centering
	\subfloat[BF Gains using oneshot and KF for frequency synch.  The standard dev. is shown as error bars.]{\includegraphics{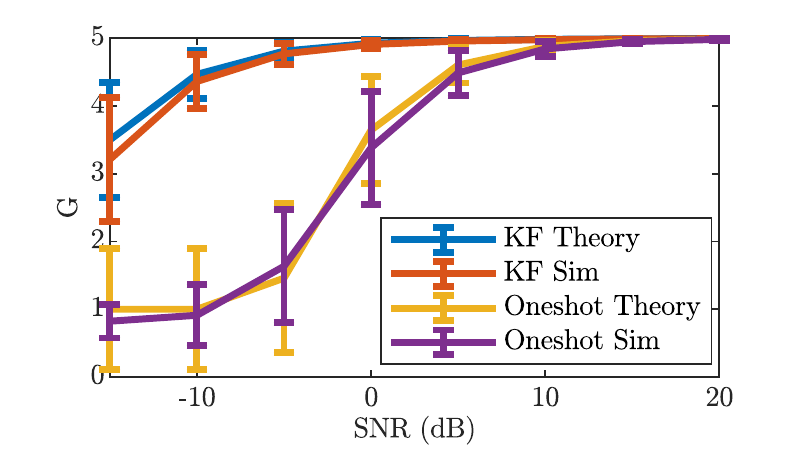}\label{fig:sim_snr_kf}}
	\ifdefined \singleCol\else	\\	\fi
	\subfloat[Phase error variance breakdown with respect to protocol stages.]{\includegraphics{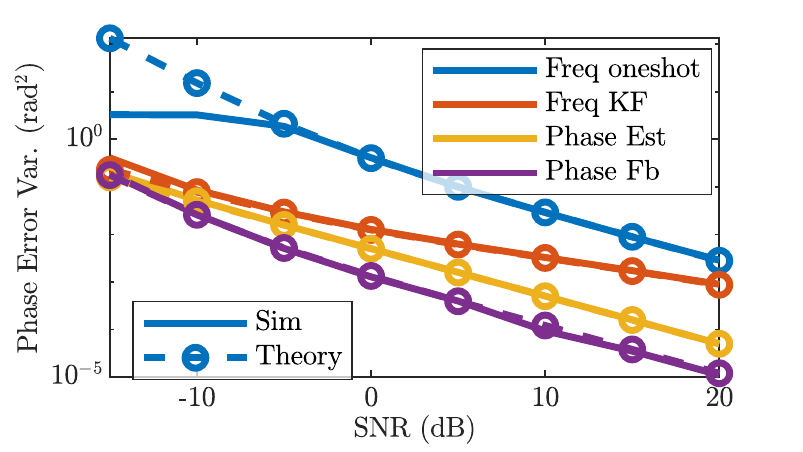}	\label{fig:sim_snr_breakdown}}
	\caption{Simulated BF Gains and phase errors  at different SNRs for $N=5$ using the waveform from Table~\ref{tbl:timing}.}
	\label{fig:sim_snr}
\end{figure}

We start by discussing the results obtained when using the oneshot  frequency estimation. The average BF gain obtained from simulations is plotted in Fig.~\ref{fig:sim_snr_kf} with the error bars representing its standard deviation. For the oneshot results, the theoretical value is obtained by calculating the variance of each estimator using (\ref{eq:freq_est_var}), (\ref{eq:ph_est_var}), and (\ref{eq:fb_est_var}), calculating $\mvarPe$ using (\ref{eq:var_pe}), then the BF gain mean and variance using Proposition~\ref{prop:G_mean_var}. From that Figure, we can see that  the theoretical  mean matches the simulations to a large extent.  As for the variances,  they match except for   SNRs below 0dB.  By plotting a breakdown of the phase error for the slave $n=3$ using (\ref{eq:var_pe}) in Fig.~\ref{fig:sim_snr_breakdown}, we see that   at SNRs below 0dB the theoretical oneshot frequency  variance is overestimated. This happened because the phase error becomes uniform and the Gaussian assumption no longer holds leading to the discrepancy in Fig.~\ref{fig:sim_snr_kf}. At these low SNRs, the BF gains are negligible and this is not a useful BF design.
From  Fig.~\ref{fig:sim_snr_breakdown},  since the phase error from the  frequency estimation error is dominant,  it would be beneficial to allocate more time to frequency estimation or use the KF to reduce its variance. %

Next, we discuss the BF results when using KF using the same Figures~\ref{fig:sim_snr_kf} and ~\ref{fig:sim_snr_breakdown}. The theoretical KF   variance is calculated using  (\ref{eq:kf_op_var})  with the measurement variance $r$ being the oneshot variance and $q$ perfectly known.  From Fig.~\ref{fig:sim_snr_kf}, we can see that both theoretical and simulated curves overlap. A small discrepancy exists at low SNR, which we attribute to an insufficient number of BF cycles. Since KF is a recursive filter, its output depends on all previous cycles and convergence is slower for high measurement noise variance~\cite{cao_exponential_2003}. Compared to the oneshot BF, at low SNR, KF provides significant BF gain improvements by reducing the frequency estimation variance and the resulting phase errors as shown in Fig.~\ref{fig:sim_snr_breakdown}. From that Figure, we  also see that as the SNR (above 0dB) becomes larger, the gap between oneshot and KF decreases. This happens because as $r$ decreases at high SNR, the ratio $r/q$ becomes small and the benefit from using KF decreases.

\subsection{Experimental Validation}
The proposed BF protocol was implemented using three  USRP B205-mini software-defined radios (SDR); two were used as BF radios and one as the destination radio. The destination radio initiates a BF cycle by transmitting the frequency synchronization preamble. The BF radios are always running the autocorrelation given by~(\ref{eq:freq_est_statistic}) and using its output power level to detect the preamble. Once detected, the frequency offset is estimated (using oneshot or KF) and corrected. Each BF radio transmits the phase estimation preamble in a preassigned time slot. The destination radio estimates the phase and feeds it back to the BF radios using the same previously discussed waveforms and estimators. Once the feedback is obtained, the radios transmit a known payload, which is received and  stored by the destination. The payload consists of three parts; each of the two BF radio transmits individually at first, then both BF radios transmit simultaneously. The magnitude of each part of the payload is estimated by averaging, then the BF gain is calculated by dividing the power of the simultaneous transmission by the sum of the individual transmissions as per~(\ref{eq:g_channel}). All the signal processing was implemented using GNURadio~\cite{noauthor_gnu_nodate} and timed burst transmissions were used for the different  stages of the protocol. The destination processing was performed on a laptop  and the BF radios on  ODROID XU4 single board computers (SBC). We conducted the experiments in the lab and on UAVs at a frequency of 915MHz with a sampling period of $\mTs=1\mu s$. %

\begin{figure}[t!]
	\centering
	\subfloat[BF gains using oneshot frequency estimation.]{\includegraphics{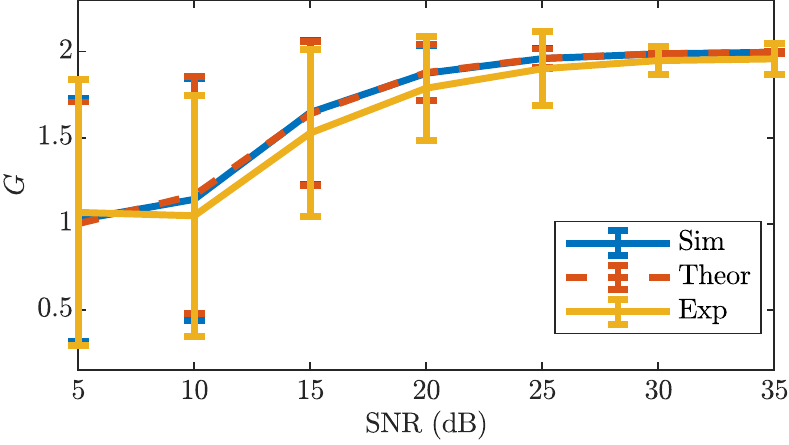}\label{fig:exp_ground}}
	\ifdefined \singleCol\else	\\	\fi
	\subfloat[BF gains using KF for frequency estimation.]{\includegraphics{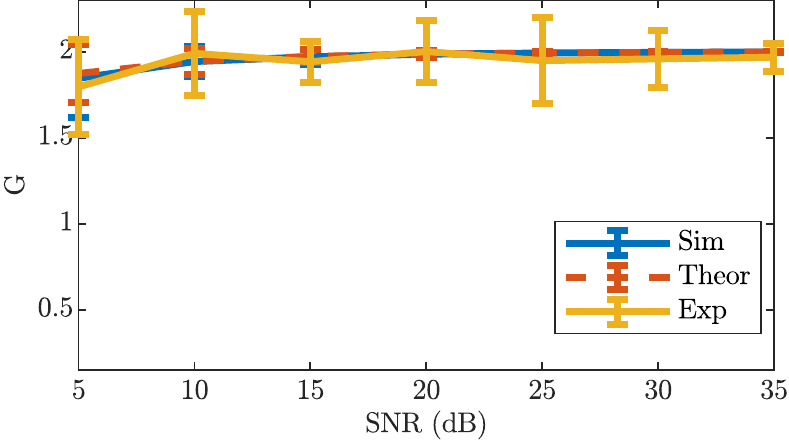}\label{fig:exp_kf}}
	\caption{Experimental results collected using $N=2$ BF software-defined-radios in a lab along with the theoretical results predicted by the BF framework and simulations.}
	\label{fig:exp_lab}
\end{figure}

\subsubsection{Lab Experiment}
We started by verifying our simulations in a lab environment with a favorable channel. The beamforming slaves were placed in proximity from each other, 2.5 meters away from the destination in an undisturbed line-of-sight environment with a measured coherence time of $0.3$s and $q=0.18$. Both the destination and BF radios were set to use the same transmit gain, which was varied in increments of 5dB to obtain different SNRs. At each SNR, 900 beamforming cycles were performed.  The timing  of the protocol is shown in Table~\ref{tbl:timing}. Notice that the guard times are much longer than in the simulations to allow the  BF signal processing to operate in real-time, which makes $\mte$ larger in~(\ref{eq:phase_error}), and thus increases $\mvarPe$  and  degrades the BF gains. 

The experimental results along with its simulated and theoretical  equivalents are shown in Fig.~\ref{fig:exp_ground}. We can see that the measured results are close to the simulation and theoretical results, which overlap.  The improvement from using KF follows  a similar trend to what was observed in Fig.~\ref{fig:exp_kf}. This result experimentally verifies our simulation setup and analysis.

\begin{figure}[t!]
	\centering
	\includegraphics{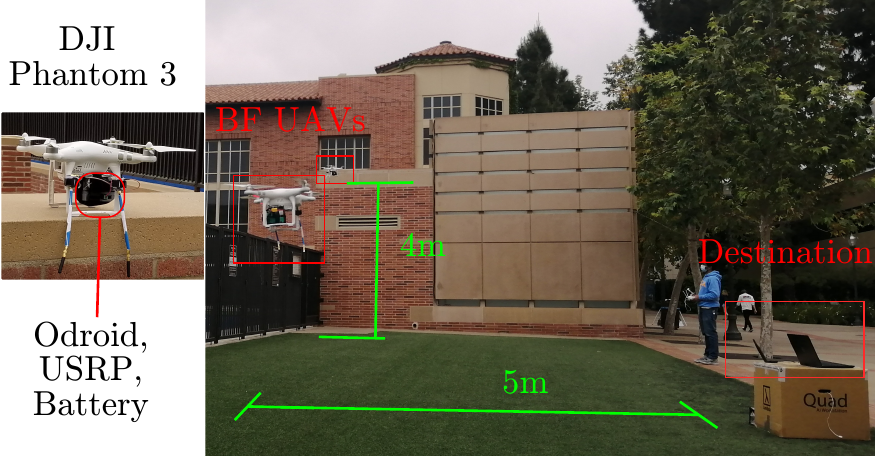}
	\caption{ The UAV experiment consists of 2 BF UAVs with SDRs mounted on-board. The UAVs were hovering freely and were not  attached to the ground by wires.}	
	\label{fig:uav}
\end{figure}

\begin{table}[t]
	\renewcommand{\arraystretch}{1.5}
	\caption{ BF UAV Results \label{tbl:uav_results}}
	\centering
	\begin{tabular}{|c|c|c|c|c|}
		\hline
		Setup & Freq &SNR (dB) & G Mean& G Stdev  \\	\hline
		Ground &  KF & 26.9 & 1.825 & 0.319 \\	\hline
		Flying & oneshot& 23.7 & 1.636 & 0.525\\	\hline
		Flying & KF& 24.9 & 1.632 & 0.438\\	\hline
	\end{tabular} 
\end{table}
\subsubsection{UAV Experiment}
Next, we move our setup from the lab to UAVs. The BF radios, consisting of the SBC and USRPs along with a battery, were mounted on two DJI Phantom 3 drones as shown in Fig.~\ref{fig:uav}. The destination radio was placed on the ground about 5m away from the UAVs which flew at a height of about 4m. The wind speed at the day of the experiment was 15Km/hr. Due to the wind and the noise of the UAV sensors, the UAVs were not stable and drifted within about a meter. The UAV operators  frequently intervened to stabilize them. 

Based on  channel estimation performed before  the experiment, the coherence time was estimated to be about $\mcoht=85ms$. Thus, the lab experiment BF cycle ($\mtcycle=180ms$) is too long for the UAV channel. For the BF  to work from the UAVs, the BF cycle was redesigned to have shorter guard times and a 10 times shorter payload as detailed in Table~\ref{tbl:timing}, yielding a reduced $\mtcycle=75ms$, which is shorter than $\mcoht$ but only with a small margin. 
The experiment was performed with three settings: 1) UAVs were on the ground and used KF for frequency synchronization, 2) UAVs were flying  and used oneshot for frequency synchronization and 3) UAVs were flying and used KF. The BF results are shown in Table~\ref{tbl:uav_results} along with the average SNR. The BF UAVs  attained about 80\% of the ideal BF gains despite the low coherence time channel. These gains are lower than the ground scenario as expected because of the shorter coherence time. As for the comparison between KF and oneshot, there is no significant difference because $r/q$ is small; $r$ is small because of the high SNR and $q$ is large because of the short coherence time.

\subsection{Emulation}
To overcome the large delays of the BF implementation and have a fair comparison between KF and Oneshot, we emulated BF over a channel trace. The channel trace was obtained by capturing a repeating ZC sequence from a flying UAV over a period of 100s. Using this trace, we emulated the BF protocol as follows; we used a duration $\mtsyn$ to estimate the frequency offset and corrected for it, then we estimated the phase offset after a delay equivalent to the protocol ($\mtgI + N \mtph + \mtgII$) and corrected for it. The feedback stage was not emulated and was assumed to be ideal. At the evaluation time $\mte$, we estimated the phase error $\mpe{}$ which for a static channel and perfect estimation should equal zero.  The variance of $\mpe{}$ calculated by emulation over the entire trace provides an estimate of $\mvarPe$ if BF was applied in this channel.  Note that  the channel trace was  collected over one capture  with a USRP operating in half-duplex. Hence, the emulation over that trace does not capture distortions due to burst transmissions and having both transmit and receive chains powered on simultaneously in the protocol implementation.  

The measured phase errors are reported in Table~\ref{tbl:emulation_results}. The first row emulates the timing used in the UAV experiment and using the  value of $\mcoht=85ms$, which is the true one, to calculate the KF $q$. The calculated phase error variance $\mvarPe$ is shown for KF and oneshot, and the theoretically  predicted mean BF gain $G$  using (\ref{eq:bf_gain_mn}) and $N=2$. Due to  the more favorable  half-duplex capture  and the ideal feedback, the predicted emulation BF gains ($\approx 1.7$) are  better than the measured ones ($\approx 1.6$). For the relatively long BF packets at a the high SNR of the capture, the predicted BF gains using both oneshot and KF are very close (1.725 abd 1.723) similar to our experimental results.
Yet the BF gains are still below 2 due to the long BF cycle, so in row 2, we emulate the protocol using a shorter cycle of $18ms$ by scaling down  $\mte$ and the phase delay. Using this shorter cycle, the BF gains increase  significantly for both KF and oneshot and approach the ideal gain of 2. This is the result  we would expect using an optimized implementation of the BF protocol having shorter guard times. Due to the high SNR, both KF and oneshot still give a similar performance. Then in row 3, we added Gaussian noise to the channel trace to make its SNR drop to 0dB. The expected BF gains for KF become significantly better than those from oneshot. This result shows that if $\mtcycle<<\mcoht$,  KF can attain significantly higher BF gains than oneshot for  distantly deployed UAVs. 
\begin{table}[t]
	\renewcommand{\arraystretch}{1.5}
	\caption{ BF Emulation over UAV Channel Trace \label{tbl:emulation_results}}
	\centering
	\begin{tabular}{|c|c|c|c|c|c|}
		\hline
		\# & SNR & $\mtcycle$ & $\mcoht$   & KF $\mvarPe$ [G]  & Onesh. $\mvarPe$ [G] \\	\hline
		1 & 24dB & 75ms & 85ms & 0.569 [1.725]& 0.567  [1.723]\\	\hline
		2 & 24dB &  18ms & 85ms& 0.11 [1.99] & 0.141 [1.98]\\	\hline
		3 & 0dB &  18ms & 85ms& 0.573 [1.85] & 0.4 [1.72]\\	\hline
	\end{tabular} 
\end{table}

\section{Beamforming System Design}
\label{sec:bf_design}
After verifying the framework, we discuss how it can be used in designing  BF systems. To design a reliable BF system, we need to specify the number of BF radios $N$ and the  duration of the preambles  to exceed a minimum post-BF SNR with a given probability. The design procedure is over two steps; first we determine $N$ and $\mvarPe$ that meet the requirements and then we design the preambles' lengths to realize $\mvarPe$ at a given pre-BF SNR.  Later in Section~\ref{sec:scenarios}, we apply the proposed design procedures for specific scenarios.

\subsection{Specifying $N$ and $\mvarPe$}
Although the pre-BF SNR ($\mSNRpre$)  is assumed constant, due to the phase error variance, the post BF SNR ($\mSNRpost$) varies randomly. A reliable BF system  has  to exceed a specified outage probability $\mpout$ such that
$
	P\left(\mSNRpost<\mSNRmin\right)\leq \mpout
$,
where  $\mSNRmin$ is the minimum SNR.
Using the gamma approximation of the BF gain distribution   and the post-BF SNR definition (\ref{eq:postBF_SNR}), we can rewrite
$
P\left(\mSNRpost<\mSNRmin\right)= 1-	F_{\mgammarv}\left(N- \frac{\mSNRmin}{\mSNRpre N}\right)
$
where $F_{\mgammarv}(x)$ is the CDF of the Gamma distribution from Proposition~\ref{prop:g_gamma} whose mean and variance depend on $N$ and $\mvarPe$. Hence,  our objective is to determine $N$ and  $\mvarPe$ which satisfy
\begin{equation}
	\label{eq:bf_requirement}
	F_{\mgammarv}\left(N- \frac{\mSNRmin}{\mSNRpre N}\right)\leq 1 - \mpout 
\end{equation}

We know the distribution of $\mgammarv$ and how $N$ and $\mvarPe$ affect it,  however, inverting (\ref{eq:bf_requirement})  to obtain an explicit relation between $N$ and $\mvarPe$ is intractable. The fact that $\mvarPe$ can depend on $N$ under fixed BF overheads further complicates analytical solutions. %
 It is easy, however, to check whether a given choice of $N$ and the corresponding $\mvarPe$ satisfies the requirements given by~(\ref{eq:bf_requirement}). Thus, we resort to numerical trial-and-error methods to find $N$ and $\mvarPe$ satisfying the requirements. %
The exact method depends on the scenario and whether $N$ is fixed or not, and thus its discussion is deferred to Section~\ref{sec:scenarios} where example scenarios are presented.

\newcommand{\mphTrgt}{\delta_{\mvarPe}} 
\newcommand{\mnTrgt}{\delta_{\mnov}} 
\subsection{Beamforming Signals Design}
For given values of $N$, $\mSNRpre$, and $\mSNRms$, we want to optimize  the time allocated to each  preamble for $\mvarPe$ to meet the system requirements.  We  identify two problems of interest; the first one is to minimize $\mvarPe$ for  limited   BF overheads and the second problem is  to minimize the  BF overheads $\mnov$ to meet a maximum allowable phase error variance. The first problem is suitable for short coherence time channels, where the BF overheads are constrained to allow time for communication within the coherence time. The second problem, on the other hand, is suited for relatively large coherence time channels, where large BF overheads are possible.  An example of each problem is provided later in Section~\ref{sec:scenarios}. %

Next, we formulate both problems. The total overheads in samples defined in (\ref{eq:overhead_n}) can be written as a function of the duration of each stage $\mnov(\mnsyn,\mnph,\mnfb)$. For fixed $N$, $\mSNRpre$ and $\mSNRms$, the phase variance  $\mvarPe$ becomes a function of the number of samples allocated to each stage $\mvarPe(\mnsyn,\mnph,\mnfb)$ defined as
\begin{equation}
	\mvarPe(\mnsyn,\mnph,\mnfb)=  (2\pi \mte)^2 \mvarF +  \mvarPh + \mvarFb
\end{equation}
 The values of $\mvarF$, $\mvarPh$, and $\mvarFb$ are dependent on the choice of estimators and are a function of $\mnsyn$, $\mnph$, and $\mnfb$ respectively.
 For our choice of estimators $\mvarPh=\mvarPhe$ defined by (\ref{eq:ph_est_var}), and $\mvarFb=\mvarFbe$ defined by (\ref{eq:fb_est_var}). As for the frequency error variance, if we use oneshot estimation $\mvarF=\mvarFe$ as defined by (\ref{eq:freq_est_var}) and if  we use the KF  $\mvarF=\mvarFk$ as defined by (\ref{eq:kf_op_var}) with $r=\mvarFe$. 

 Note that  for a chosen Zadoff-Chu sequence of a  length $\mlzc$,  we can only optimize the number of repetitions $\mnzc$ to change $\mnsyn$. Hence, for fixed $N$, the problem P1 can be written as
 \begin{align}
 P1:	& \underset{ \mnzc,\mnph,\mnfb }{\text{minimize}} & &  \mvarPe(\mnzc \mlzc,\mnph,\mnfb)  \\
 & \text{subject to} & &   \mnov(\mnzc \mlzc,\mnph,\mnfb)\leq  \mnTrgt 
 \ifdefined \singleCol , \ \ \else \nonumber\\ \myamp \myamp \myamp \fi
 \mnzc,\mnph,\mnfb \in \Z^+, \ \ \mnzc\geq 2 
 \end{align}
 where  $\mnTrgt$ is maximum  overhead length which depends on the channel coherence time,  and $\Z^+$ is the set of positive integers.  For a maximum allowable phase error $\mphTrgt$, the second problem P2 can be written as
\begin{align}
P2:	& \underset{ \mnzc,\mnph,\mnfb }{\text{minimize}} & & \mnov(\mnzc \mlzc,\mnph,\mnfb) \\
	& \text{subject to} & & \mvarPe(\mnzc \mlzc,\mnph,\mnfb) \leq  \mphTrgt
	\ifdefined \singleCol , \ \ \else \nonumber \\ \myamp \myamp \myamp \fi
	 \mnzc,\mnph,\mnfb \in \Z^+, \ \ \mnzc\geq 2 
\end{align}

Then, we argue that for our choice of estimators, both problems are convex  with respect to their variables and thus are easy to solve. Except for the KF, all these estimators take the form $f(x)=\frac{c_1}{x}+ \frac{c_2}{x^2}$ with respect to their variables for some positive $c_1$ and $c_2$ where $x$ is strictly positive, hence they are all convex over their domain. As for the KF, when substituting for $r$, it takes the form $f(x)=c_1 + \sqrt{c_2 + \frac{c_3}{x}+ \frac{c_4}{x^2}} $  with respect to its positive variable $x$ for some positive $c_1,c_2,c_3$ and $c_4$. This can be rewritten as $f(x)=c_1 + \| \b{y} \| $ where $ \b{y} = [\sqrt{c_2} , \frac{\sqrt{c_3}}{\sqrt{x}} , \frac{\sqrt{c_4}}{x}]^T$. The norm is convex and non decreasing and $\frac{\sqrt{c_3}}{\sqrt{x}}$ and $\frac{\sqrt{c_4}}{x}$ are convex for positive $x$. By applying the  composition rule~\cite{boyd_convex_2004}, the KF variance is convex. Hence,  $\mvarPe(\mnzc \mlzc,\mnph,\mnfb) $ is convex with respect to its arguments for all of our estimators. As for $\mnov(\mnzc \mlzc,\mnph,\mnfb)$, it is an affine combination of its arguments. This makes both problems P1 and P2  integer convex problems, which can be optimally solved using CVX with a mixed integer solver~\cite{grant_cvx_2014}.

\newcommand{\mGreq}{G_{\text{req}}}
\newcommand{\mNmin}{N_{\text{lb}}}
\newcommand{\mNmax}{N_{\text{ub}}}
\section{Beamforming Design Scenarios}
The proposed BF framework and the derived relations can be applied to many BF scenarios. In this ection, we discuss the design procedures for two example scenarios. 
In the first example, we consider a large swarm of small UAVs; we want to determine the minimum $N$ to satisfy the SNR requirements. Due to the UAVs' high mobility, the channel coherence time is small and the BF overheads are constrained. This example maps to the problem P1. In the second example, we consider $N=4$ weather balloons sending short payloads. Due to the long coherence time resulting from the balloons slow motion,  large BF overheads are possible. However, to avoid energy wasted on unneeded transmission,  our objective is to minimize the BF overheads while satisfying the SNR requirement. This example maps to the problem P2. 
\label{sec:scenarios}
\subsection{Swarm of Small UAVs}
\label{sec:scenarios_small_uav}
\begin{figure*}[t!]
	\centering
	\subfloat[Number of BF Radios obtained  assuming ideal BF   ($N_{lb}$) and obtained using our framework ($N$).]{\includegraphics[width=0.32\textwidth]{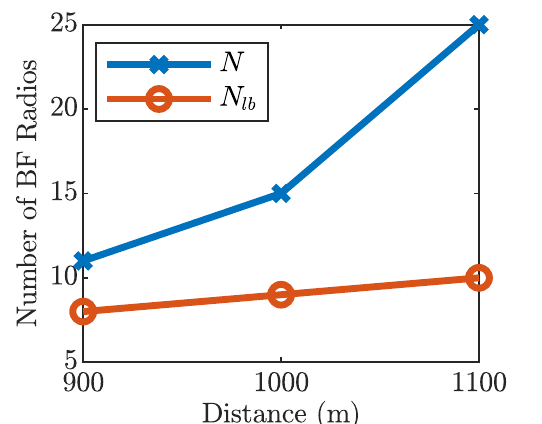}\label{fig:small_uav_n}} \hspace{1mm}
	\subfloat[CDF of simulated BF gains  using  $N$ (from our framework). The requirement given by $\mSNRmin$ and $\mpout$ is satisfied.]{\includegraphics[width=0.32\textwidth]{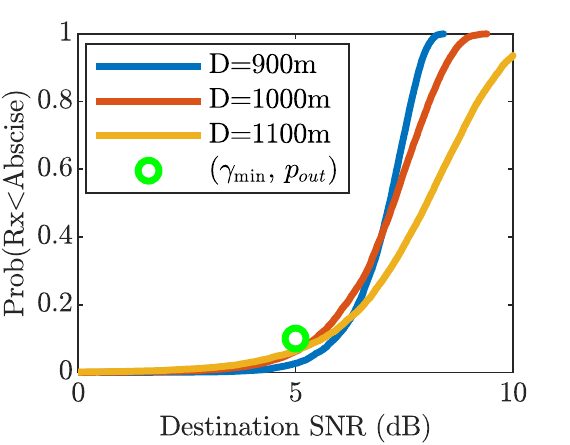}\label{fig:small_uav_cdf_calc}}\hspace{1mm}
	\subfloat[CDF of simulated BF gains using   $\mNmin$ (assuming ideal BF). The requirement given by $\mSNRmin$ and $\mpout$ is violated.]{\includegraphics[width=0.32\textwidth]{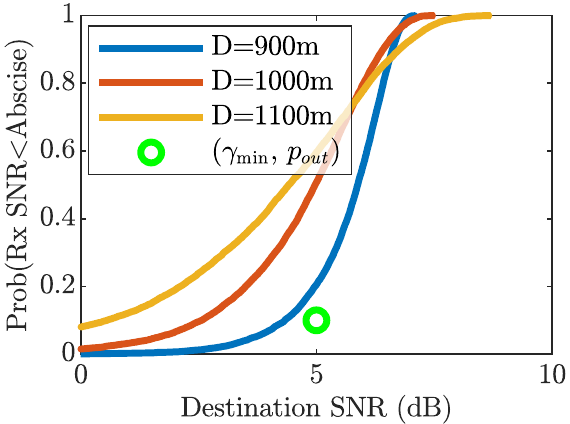}\label{fig:small_uav_cdf_ideal}}
	\caption{Results for minimizing $N$ in a  swarm of small UAVs assuming a fixed BF overhead.  Using $N$ obtained from our approach, the SNR requirement is satisfied as verified by simulations.}
	\label{fig:small_uav}
\end{figure*}

A swarm of $\mNmax$ small UAVs is deployed in an urban environment  for an application like crowd monitoring~\cite{trotta_persistent_2020}.  A large number of small UAVs is deployed and they continuously transmit data. To avoid a large overhead in data sharing among UAVs for BF,  we want to determine the minimum  number of UAVs to beamform such that the destination SNR exceeds a minimum of $\mSNRmin=5dB$ for 90\% of the time ($\mpout=0.1$).

For the urban channel, we consider a channel having a path loss coefficient of 3.7~\cite{goldsmith_wireless_2005} and a coherence time of 10ms~\cite{chakkour_coherence_2018}. The maximum transmit power of each UAV is $P_T=0dBm$ and of the destination $\mPdt=20dBm$. Communication takes place over a frequency of 915MHz using  a sampling time of $\mTs=1\mu  s$ and a BW of 1MHz   and all radios  have a noise figure of 3dB. By performing the link budget calculation, the SNR from an individual UAV at 1Km is close to -13dB, so  the minimum required BF gain is $\mGreq=18$dB. Assuming ideal BF gain of $N^2$, the required gain can be achieved using only 8 BF radios. However,  due to the short coherence time, the entire BF packet is assumed to be limited to 5ms and based on the payload required by the application only 1ms of BF overhead is allowed. At this low SNR and with this short BF overhead, the ideal beamforming gains are not achievable and large BF variance is expected. We need to use more than 8 BF radios so that the  SNR exceeds 5dB for 90\% of the time as required. Our objective is to determine the minimum $N$ and the duration of each preamble.

We use our analytical framework to find the minimum $N$. Since for fixed overheads $\mnov$, $\mvarPe$ depends on $N$, we need to solve P1 to calculate  $\mvarPe$ for each $N$. The proposed approach is summarized in~Algorithm~\ref{alg:small_uav} and it works as follows;
we  start from the lower bound on $N$, which occurs when assuming ideal BF  $\mNmin=\left\lceil \sqrt{\mGreq} \right\rceil$   and increment $N$ until the requirement is satisfied. For each $N$,  we solve the minimum phase error problem $P1$ to obtain $\mvarPe$. Using the resulting $\mvarPe$,   we substitute in (\ref{eq:bf_requirement}) to determine if the requirement is satisfied or not. The first $N$ satisfying the requirement is the minimum $N$ meeting the SNR requirements. If the maximum number of available BF radios $\mNmax$ was reached without satisfying (\ref{eq:bf_requirement}), another approach needs to be considered to meet the requirements like increasing  the BF overhead  or the transmit power of the radios.  Since the BF is performed periodically and $\mtcycle<\mcoht$, we assume that KF is used for frequency tracking.

The calculated $N$ for different distances is shown in Fig.~\ref{fig:small_uav_n} along with $\mNmin$ calculated assuming ideal BF gain.  To verify that the obtained solution meets our design criteria, we simulated 10K BF cycles of the BF protocol using using the calculated $N$ and the optimized waveforms obtained from $P1$ at each distance. The destination SNR was measured and its empirical CDF for the proposed $N$  and the ideal $\mNmin$  are plotted in Fig.~\ref{fig:small_uav_cdf_calc} and~\ref{fig:small_uav_cdf_ideal} respectively. From these Figures, we see that the required outage probability is met using the proposed $N$. Thus, our problem solution and the underlying analysis can be used to  design  reliable BF systems satisfying the design requirements as verified by simulations. On the other hand, relying on the ideal $\mNmin$  is expected to yield lower BF gains than the desired ones in realistic deployment scenarios.
   \begin{algorithm}[t!]
   	\SetAlgoLined
   	\SetKwInOut{Input}{input}
   	\SetKwInOut{Output}{output}
   	\Input{$\mNmin,\mNmax,\mnov,\mpout,\mSNRmin$}
   	\Output{Solved, $N$, BF waveform}
	Solved := False \;
   	\For{\upshape  $n_i$ = $\mNmin$ to $\mNmax$}{
   			Solve P1 to determine $\mvarPe$\;
   			\If{ $n_i$ and $\mvarPe$ satisfy (\ref{eq:bf_requirement})}
   			{Set Solved to True, $N$ to $n_i$, and BF waveform to solution of P1, and  exit \;} 
   	}
   	\caption{} \label{alg:small_uav}
   \end{algorithm}

\subsection{Weather Balloons}
\begin{figure}[t!]
	\centering
	\subfloat[The minimum BF overhead  to satisfy the SNR requirement.]{\includegraphics{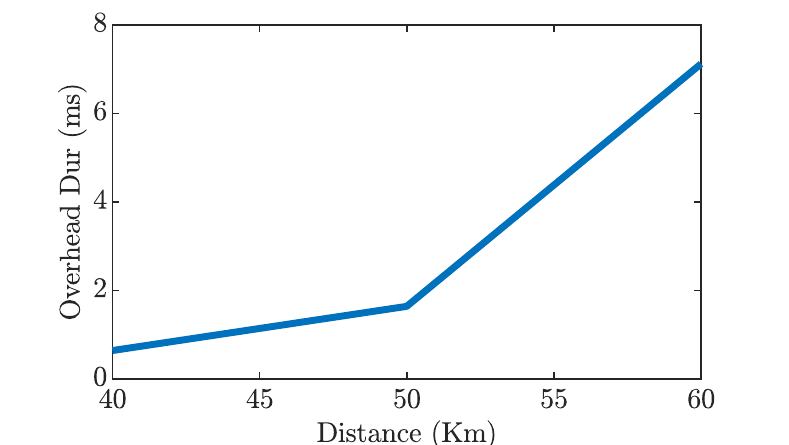}\label{fig:weather_balloon_n}}
	\ifdefined \singleCol\else	\\	\fi
	\subfloat[CDF of simulated BF gains  using the minimum overhead. The SNR requirements given by $\mSNRmin$ and $\mpout$ are met.]{\includegraphics{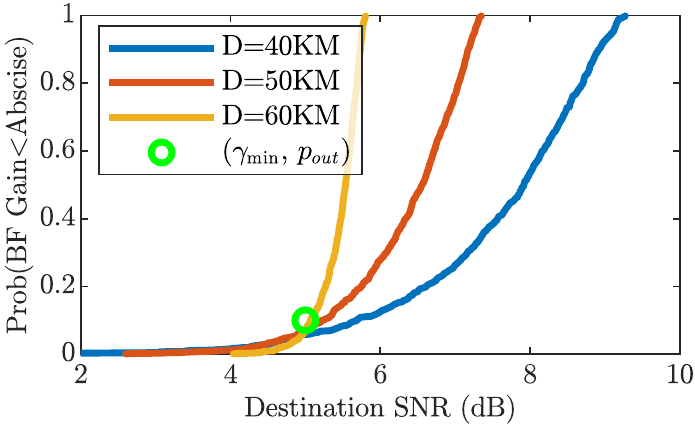}\label{fig:weather_balloon_cdf_calc}}
	\caption{For   $N=4$ weather balloons, in a long coherence time channel, the minimum overheads obtained using our framework satisfy the SNR requirements.}
	\label{fig:weather_balloon}
\end{figure}

Weather balloons are deployed at high altitudes to perform atmospheric measurements and report them back to the ground. We consider $N=4$ weather balloons deployed at a distance of 50KM from the destination radio. Due to their high altitude, the channel is dominated by line-of-sight propagation and we consider a path loss coefficient of $2$ and a large channel coherence time exceeding $100$ms. The large channel coherence time allows for much longer BF overheads. However, to economize the balloon payload battery power, we want to minimize the transmission time.  Our objective is to determine the smallest BF overheads to attain  a  received SNR exceeding a minimum of $\mSNRmin=5dB$ for 90\% of the time ($\mpout=0.1$).
 We use the same  power, frequency, bandwidth,  and noise parameters as the previous scenario except $P_T=10$dBm is larger. The SNR from a single radio is -4.6dB and thus the required BF gain at 50KM distance is 9.6dB. Assuming that the measurements are infrequent and not periodic, we use oneshot  frequency estimation. 
 
 To design this system, we find the minimum phase error needed to satisfy the requirement ($\mphTrgt$), then we find the shortest overhead to meet this phase error.
 Since, for fixed $N$, increasing $\mvarPe$ decreases the average BF gain and vice versa, we determine $\mphTrgt$ by applying the bisection method on (\ref{eq:bf_requirement}). Then, using $\mphTrgt$, we solve the problem $P2$ to determine the minimum overhead. If the minimum overhead  makes the BF packet  exceed the channel coherence time, the solution is not valid and we need to consider another alternative like increasing the transmit power. The minimum overheads obtained are shown in Fig.~\ref{fig:weather_balloon_n} for different distances. Then, we simulated the BF protocol at these SNRs using the waveforms obtained from P2 and plotted the empirical CDF of the destination SNR in Fig.~\ref{fig:weather_balloon_cdf_calc}. We can see that the proposed solution approach  meets the required outage probability, which verifies the solution and all the underlying analysis.

\section{Conclusion}
In this work, we developed and verified a mathematical framework to model the BF performance for a destination-led BF protocol. The BF gains distribution was approximated by a gamma distribution assuming  a zero mean normally distributed combining phase errors and the proposed distribution was verified using simulations. The effect of the pre-BF SNR and the preamble lengths on the combining phase error was derived for our choice of estimators.  Using software-defined radios, in a lab, we experimentally verified the  predictions of our BF framework. The BF radios were mounted on UAVs and were shown to exceed $80\%$ of the ideal BF gains despite the low coherence time channel.  %
 The proposed framework can be used to design BF systems for a given deployment as illustrated by two example scenarios.

 Even though we only considered a specific BF protocol  and  only  two example scenarios, the proposed framework can support many protocol variations and use cases. For the  protocol, the framework is  applicable for any other choice of estimators as long as their  phase variance can be expressed mathematically.   As for the scenarios,   heuristics can easily be developed to optimize over a combination of the SNR, preamble lengths, and the number of BF slaves, enabling the framework to adapt to many different deployment scenarios.

\appendix
\subsection{Proof of Proposition~\ref{prop:kf_var}}
\label{ap:kf_var}
The variance error of the KF output is given by $p_{k|k}$ and we want to calculate its value. Substituting~(\ref{eq:kf_gain}) into (\ref{eq:kf_var_udpate}), we get
$
	p_{k+1|k}	=\frac{rp_{k|k-1}}{p_{k|k-1}+r}+q
$.
At steady state $p_{k+1|k}=p$ for all $k$ and we get a simple form of the algebraic Riccati equation
$
	p	=\frac{rp}{p+r}+q
$
~\cite{bertsekas_dynamic_2000}.
Solving the  equation, we get $p_{k+1|k}=p=\frac{q+q\sqrt{1+4\frac{r}{q}}}{2}$. Using (\ref{eq:kf_var_udpate}), we get $\mvarFk=p_{k|k}=\frac{-q+q\sqrt{1+4\frac{r}{q}}}{2}$.
\subsection{Proof of Proposition~\ref{prop:G_mean_var}}
\label{ap:G_mean_var}
\begin{align}
	G &= \frac{1}{N}	\left|\sum_{n=1}^{N}e^{j\mpe{n}}\right|^{2} 	= \frac{1}{N} \sum_{n=1}^{N}e^{j\mpe{n}}\sum_{m=1}^{N}e^{-j\mpe{m}} 	
	\ifdefined \singleCol \else \nonumber \\	\myamp   \fi
	=1+\frac{2}{N}\sum_{m=1}^{N}\sum_{n=m+1}^{N}\cos(\mpe{n}-\mpe{m}) \label{eq:g_simple}
\end{align}
Using the fact that for a zero mean Gaussian RV $x$, $\E{\cos{x}}=e^{-\mvar{x}/2}$ \cite{richard_brown_receiver-coordinated_2012},  we get
\begin{align}
	\E{G} %
	&= 1+(N-1)e^{-\mvarPe}
\end{align}
\begin{align}
	\mvar{G}	&=\frac{4}{N^2} \mvar{\sum_{m=1}^{N}\sum_{i=m+1}^{N}\cos(\mpe{i}-\mpe{m})} \\ %
	&\hspace{-0.5em}=\frac{4}{N^2}\sum_{m=1}^{N}\sum_{i=m+1}^{N}\mvar{cos(\mpe{i}-\mpe{m})}
\ifdefined \singleCol \else \nonumber\\	\myamp \hspace{-0.5em}  \fi
+\frac{8}{N^2}\sum_{m=1}^{N}\sum_{i=m+1}^{N}\sum_{p=i+1}^{N}\mcov{\cos(\mpe{i}-\mpe{m}),\cos(\mpe{m}-\mpe{p})} \label{eq:gvar_deriv_1}\\
	&= \frac{(N-1)}{N}(e^{-\mvarPe}-1)^{2}\left((e^{-\mvarPe}-1)^{2}+2Ne^{-\mvarPe})\right)\label{eq:gvar_deriv_2}
\end{align}
where $\mcov{x,y}$ denotes the covariance of RVs $x$, $y$. Line (\ref{eq:gvar_deriv_1}) was obtained using the fact that  $\mvar{\sum_{m=1}^{M} x_m}=\sum_{m=1}^{M}\mvar{x_m}+2\sum_{m=1}^{M} \sum_{n=m+1}^{N} \mcov{x_m,x_n}$ for any correlated $M$ RVs $x_m$ and by simplifying the summations. Line (\ref{eq:gvar_deriv_2}) uses the fact that for a zero mean Gaussian RV $\mvar{\cos x}=\frac{1}{2}(e^{-\mvar{x}-1})^2$~\cite{richard_brown_receiver-coordinated_2012} and using that  $\mcov{\cos(\mpe{i}-\mpe{m}),\cos(\mpe{m}-\mpe{p})} =0.5e^{-3\mvarPe}+0.5e^{-\mvarPe}-e^{-2\mvarPe} $ as can be shown using the Gaussian RV relations from~\cite{richard_brown_receiver-coordinated_2012}, the definition of covariance, and some  trigonometric identities.
\subsection{Proof of Proposition~\ref{prop:g_gaussian}}
\label{ap:prop_g_normal}
\begin{figure}[t!]
	\centering
	\includegraphics[scale=0.8]{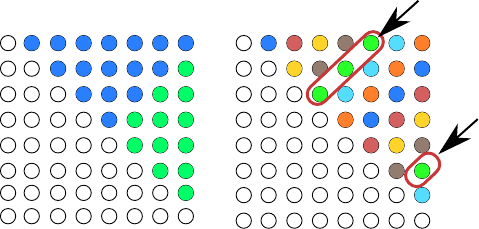}
	\caption{Summation order for the matrix $\mXX$.}	
	\label{fig:matrix}
\end{figure}

	We start be considering the simplified definition of $G$ from (\ref{eq:g_simple}). We rewrite the elements of the summation as the $N\times N$ matrix $\mXX$, such that its element $ \mX{m,n}=\frac{2}{N}\cos(\phi_{m}-\phi_{n})$. This yields
$
	G=1 +\sum_{m=1}^{N}\sum_{n=m+1}^{N}\mX{m,n}
$.
The summation is over the upper diagonal elements of the matrix. Our objective is to rewrite the inner sum as independent RVs of length proportional to $N$ to invoke the central limit theory (CLT). To achieve that, we must  avoid reusing the same value of $\mpe{n}$ in the inner sum, that is, the inner sum elements  should have unique column and row indices. 
\begin{align}
	&	\sum_{m=1}^{N}\sum_{n=m+1}^{N}\mX{m,n} = \sum_{n=2}^{N}\sum_{m=1}^{n-1} \mX{m,n} \label{eq:g_gaussian_1}\\
	&=\sum_{n=2}^{N}\sum_{m=1}^{\min(n-1,N-n+1)} \mX{m,n}+\sum_{n=N/2+1}^{N}\sum_{m=N-n+1}^{n-1} \mX{m,n} \label{eq:g_gaussian_2}\\
	&=\sum_{n=2}^{N}\left(\sum_{m=1}^{\left\lfloor n/2\right\rfloor }\mX{m,n-m+1}+ \sum_{m=1}^{\left\lfloor (N+1-n)/2\right\rfloor } \mX{N+1-(n-m+1),(N+1)-m} \right)\label{eq:g_gaussian_3}
\end{align}
The summation in~(\ref{eq:g_gaussian_1}) rewrites the equation from column wise to row wise. In Line~(\ref{eq:g_gaussian_2}), we split the elements of the summation at the upward diagonal as illustrated in the first image of Fig.~\ref{fig:matrix} for an $8\times 8$ matrix. In Line (\ref{eq:g_gaussian_3}),the inner summation is  rewritten as two summations over the upward diagonal elements as shown in different colors in the second image of  Fig.~\ref{fig:matrix}. From~(\ref{eq:g_gaussian_3}), each element of the inner summation consists of about $N/2$ terms\footnote{For odd $N$, the number of elements is either $N/2$ or $N/2-1$. This difference is insignificant for large $N$} and none of the terms have common rows or columns, thus consist of independent RVs. We can rewrite the inner sum as the RV $b_n$ as follows
\begin{equation}
	\label{eq:g_bn}
	b_n=\sum_{m=1}^{\left\lfloor n/2\right\rfloor }\mX{m,n-m+1}+ \sum_{m=1}^{\left\lfloor (N+1-n)/2\right\rfloor } \mX{N+1-(n-m+1),(N+1)-m}
\end{equation}
The variable $b_n$ consists of identical independent RVs. Hence, for large $N$, the distribution of $b_n$ converges to a Gaussian distribution. Lastly, we can rewrite $G$ as
\begin{equation}
	\label{eq:g_iid}
	G=1 + \sum_{n=2}^{N} b_n
\end{equation}
The variables $b_n$ are correlated Gaussian RVs, hence their sum is  Gaussian.  This proves that for large $N$,  $G$ is Gaussian and its mean and variance are given by Proposition~\ref{prop:G_mean_var}.
\subsection{Proof of Proposition~\ref{prop:g_gamma}}
\label{ap:g_gamma}
	We start  this proof by considering the case of small $\mvarPe$ and then discuss the case of large $N$. Since $\mph{n}$ are zero mean and assuming small $\mvarPe$, $\phi_{m}-\phi_{n}$ is typically small and we can use the Taylor expansion of cosine around zero to simplify $ \mX{m,n}$ (as defined in Appendix~\ref{ap:prop_g_normal}) as 
$
	\mX{m,n}\approx \frac{2}{N} \left( 1 - \frac{(\phi_{m}-\phi_{n})^2}{2} \right)
$.
Then, we can rewrite (\ref{eq:g_bn}) as
$
	b_n= 2\frac{s_n}{N} 
	-  \chi_n 
$ 
where $s_n=\left\lfloor n/2\right\rfloor + \left\lfloor (N+1-n)/2\right\rfloor$ is the number of elements of $b_n$ and $\chi_n =\frac{1}{N} 	\sum_{r=1}^{s_n} (\phi_{m_r}-\phi_{n_r})^2$ with $m_r$ and $n_r$ corresponding to the indexes from (\ref{eq:g_bn}).   The summation in $\chi_n$ is over independent zero mean Gaussian RVs that are squared, hence $\chi_n$ follows the Chi-squared distribution. We can rewrite $G$ as 
\begin{align}
	G&=1+ \frac{2}{N} \frac{N(N-1)}{2} - \sum_{n=2}^{N} \chi_n
	= N - \mgammarv
\end{align}
where $\mgammarv=\sum_{n=2}^{N} \chi_n$ is the sum of correlated Chi-squared RVs.  The distribution of the summation of correlated Chi-squared RVs can be obtained using the Gamma distribution~\cite{gordon_cumulative_1983}. The shape $K$ and scale $\theta$ parametrization of the resulting Gamma distribution  can be calculated to realize the mean and variance of $\mgammarv$~\cite{ferrari_note_2019}. Using the mean and variance of $G$ from Proposition~\ref{prop:G_mean_var}, we get the following equations for the mean and variance respectively
\begin{equation}
	K \theta = N-1-(N-1)e^{-\mvarPe}
\end{equation}
\begin{equation}
	K \theta^2 = \frac{(N-1)}{N}(1-e^{-\mvarPe})^{2}\left((1-e^{-\mvarPe})^{2}+2Ne^{-\mvarPe})\right)
\end{equation}
Solving these two equations, we get the values of $K$ and $\theta$ in (\ref{eq:g_gamma_k}) and (\ref{eq:g_gamma_theta}). This proof is based on the assumption that $\mvarPe$ is small.
For large values of $N$, $K$ becomes large, and the Gamma distribution converges to a Gaussian distribution  with mean $K\theta$ and variance $K\theta^2$~\cite{dasgupta_normal_2010}, which is the true distribution of $G$ as shown in Proposition~\ref{prop:g_gaussian}.

\bibliographystyle{IEEEtran}
\bibliography{references}

\end{document}